\documentclass[reprint,aip,apl,superscriptaddress,showpacs,floatfix]{revtex4-1}
\usepackage[pdfborder=0 0 0]{hyperref}
\usepackage[dvips]{graphicx}
\usepackage{amsmath}
\usepackage{color}
\usepackage{amsfonts}
\usepackage{amssymb}
\usepackage{epstopdf}
\usepackage{colortbl}
\DeclareGraphicsExtensions{.eps,.png,.pdf}
\begin{document}
\title{Spatial symmetry breaking in single-frequency CCP discharge with transverse magnetic field}
\author{Sarveshwar Sharma} 
\email[e-mail: ]{sarvesh@ipr.res.in}
\affiliation{Institute for Plasma Research, Bhat, Gandhinagar 382 428, India}
\affiliation{Homi Bhabha National Institute, Anushaktinagar, Mumbai 400 094, India}
\author{Igor Kaganovich}
\author{Alexander Khrabrov}
\affiliation{Princeton Plasma Physics Laboratory, Princeton University, Princeton, New Jersey 08543, USA} 
\author{Predhiman Kaw} 
\thanks{Deceased}
 \author{Abhijit Sen}
 \affiliation{Institute for Plasma Research, Bhat, Gandhinagar 382 428, India}
\newcommand{\beq}{\begin{equation}}
\newcommand{\eeq}{\end{equation}}
\newcommand{\beqstar}{\[}
\newcommand{\eeqstar}{\]}
\newcommand{\bea}{\begin{eqnarray}}
\newcommand{\eea}{\end{eqnarray}}
\newcommand{\beastar}{\begin{eqnarray*}}
\newcommand{\eeastar}{\end{eqnarray*}}
\begin{abstract} 
An independent control of the flux and energy of ions impacting on an object immersed in a
plasma is often desirable for many industrial processes such as microelectronics manufacturing.
We demonstrate that a simultaneous control of these quantities is possible by a suitable 
choice of a static magnetic field applied parallel to the plane electrodes in a standard single 
frequency capacitively coupled plasma device. Our particle-in-cell simulations show 
a 60\% reduction in the sheath width (that improves control of ion energy) and a 
four fold increase in the ion flux at the electrode as a consequence of the altered 
ion and electron dynamics due to the ambient magnetic field. A detailed analysis of the 
particle dynamics is presented and the optimized operating parameters of the device 
are discussed. The present technique offers a simple and attractive alternative to 
conventional dual frequency based devices that often suffer from undesirable limitations 
arising from frequency coupling and electromagnetic effects.
\end{abstract}

\maketitle

In plasma devices used for industrial processes such as etching, surface engineering, material deposition etc. the ion impact energy and the flux of ions incident on a target object are important parameters that influence the quality and throughput of the entire process. In the commonly used capacitively coupled plasma (CCP) device, typically  operated by a single radio frequency (RF) source, e.g. at $13.56~\mathrm{MHz}$ \cite{LibLicPPDM, APL_1995_67_191, APL_2008_93_251602}, these parameters are governed by the geometry, the operating pressure and the input power of the device. For a fixed geometry and pressure both the ion energy and ion flux vary with the input power. Therefore in a single frequency capacitively coupled plasma (SF-CCP) device the ion energy and ion flux cannot be controlled independently 
\cite{Lib1988, God1976, PopGod1985, Kag2002, KawLieLic2006, KagPolThe2006, ShaTun2013, ShaMisKaw2014}. 
To overcome this constraint some alternate schemes have also been developed in the past. A dual-frequency device (DF-CCP) \cite{GotoLowe, RobiBoy, BoyElling, sharmaTurner2013, APL_2008_93_071501} that is now widely used in the semiconductor industry utilizes a high frequency 
($f_h$) component to largely control the plasma density (and hence the ion flux) while a low frequency ($f_l$)
component influences the sheath width and thereby the ion energy. However, independent control of these two parameters can get compromised if the two frequencies are too close to each other because of mutual
coupling between the $f_l$ and the $f_h$ frequencies \cite{SchuGans, KawLieLic2006, TurChab, GansSchu2006}. 
One way to minimize this frequency coupling is to choose $f_h$ to be very high compared to $f_l$. 
However, for very high frequencies, say $f_h>70~\mathrm{MHz}$, electromagnetic effects 
can limit the uniformity of the reaction process \cite{PSST_2002_11_283, APL_2003_83_243, APL_2005_86_021501}. 
Other alternative schemes exploit electrical asymmetry effects \cite{Czarnetzki2011, Economou2013, Lafleur2016}.

At a fundamental level, the superiority of the DF-CCP over the SF-CCP arises from the fact that the two disparate frequencies of the former can independently influence the dynamics of the electron and ion species of the plasma. The $f_h$ has a large influence on the electrons while the $f_l$ acts on the ions. As is well known, such a difference in the dynamical behavior of electrons and ions can also be brought about by employing a static magnetic field 
of a strength such that the electrons are magnetized while the ions are not. The question is, will application of such a magnetic field parallel to the electrodes in a SF-CCP provide it with an ability to achieve simultaneous control of the ion flux and ion energy to attain the desired optimum values for a given industrial application. Our present work is devoted to exploring such a possibility by carrying out extensive particle-in-cell simulations of a model SF-CCP with an applied static magnetic field. Magnetic fields have been employed in the past by many researchers in the area of magnetically enhanced reactive ion etching (MERIE) \cite{JAP_2003_94_1436,Muller_ME_10_1989,JAP_2004_95_834,Lib_IEEETPS_19_1991,Hutchinson_IEEETPS_23_1995,park_IEEETPS_25_1997}. Kushner \cite{JAP_2003_94_1436} used a 2-D hybrid fluid simulation for an Argon plasma and reported that the performance of low-pressure CCP discharges can be improved by using a transverse (parallel to substrate) static magnetic field (tens to hundreds of Gauss) to increase the plasma density.  In Ref. \cite{You2011},  S. J. You {\it et al.} have experimentally studied the influence of a magnetic field on asymmetric SF-CCP argon discharges (operated at $13.56~\mathrm{MHz}$) at low and intermediate pressures. They observed a shift of the density along the electrodes arising from an $E \times B$ drift due to the electric field perpendicular to the
electrodes and $B$ parallel to the electrodes. In the IBM MERIE reactor the magnetic field is slowly rotated in the plane parallel to electrodes to mitigate this effect. Although the MERIE process has been studied for long, the magnetic field induced asymmetry effect has escaped attention. Other recent studies devoted to CCP operation in the presence of an external magnetic field \cite{JanShihabAbd46_2013, ZadirievRukKralVav11_2016, You2011}, have explored somewhat different effects. S. Yang {\it et al} \cite{Yang2017,Yang2018} have used an asymmetric magnetic field with variable gradients to create asymmetry in the configuration of a CCP device. Their particle simulation studies show that the magnetic field asymmetry provides a means of independently controlling the ion flux and ion energy.  They did not report any enhancement in the ion flux, or efficient control of the ion energy with a weak magnetic field.   Our present 1D-3V (i.e. one spatial and three velocity) PIC simulations show that such a capability can also be created in a symmetric SF-CCP discharge by application of a uniform $B$ field. We find that for a given RF frequency the magnitude of the central density and its location as well as the sheath width at the electrodes can be precisely controlled by the strength of the magnetic field. Therefore, by varying the magnetic field we can control both the ion density (and hence the ion flux) and the ion energy impacting on the electrode.\\

Our simulations of CCP discharge have been performed with the well tested and widely used electrostatic direct implicit code EDIPIC (details can be found in Ref.  \cite{Dmytro_2006, CamKhraKag_19_2012, SheeHershKagaWang_111_2013, CamHong_103_2013, John_PSST_26_2017}). We use a simple model discharge between two plane parallel electrodes that are separated by a gap of $10~\mathrm{cm}$ with a RF voltage of $1000~\mathrm{V}$ and frequency $27.12~\mathrm{MHz}$ applied to one of the electrodes (PE) while the other is grounded (GE); note that these designations are arbitrary for a 1d system and observed asymmetric structures can be inverted by changing the initial phase of the driving waveform. 
The choice of $L=~10~\mathrm{cm}$, the discharge gap, is completely arbitrary and not related with any commercial plasma system. However the electrode dimensions are assumed to be much larger than the gap distance so that a one dimensional spatial assumption holds good. Likewise the applied voltage is chosen to be large at 1kV keeping in mind the use of He as the working gas in our model. Compared to Argon, He has a higher ionization potential and much lower ion mass requiring a higher voltage for its breakdown and sustenance. Such voltages are not unrealistic and rf voltages in the kV range are routinely used in many present day CCP  processing reactors \cite{Godyak_PRL_68_1992,Godyak_PSST_36_1992, Godyak_PRL_65_1990}. One of the earliest examples of a high voltage ($\sim 2 kV$) He discharge is the work by Godyak {\it et al} \cite{Godyak_IEEETPS_14_1986}.   An external magnetic field ($B$), applied parallel to the electrodes, is varied from $0~\mathrm{G}$ to $70~\mathrm{G}$. Note that although the code is one dimensional in space it is three dimensional in velocity space so the {\it $E \times B$} motion of the charged particles is correctly simulated. In simulations it is easier to assume a given potential on electrodes and allow for the current form to adjust accordingly and hence the external circuit is not included in the simulation. It should be pointed out however that the external circuit has to allow a time-averaged net current because it exists in the asymmetric state. We may also mention here that in order to test the robustness of the magnetic field induced asymmetry effects we have carried out additional simulations using $L=6\;cms, B=30\;G, V=1000V$ and $L=10\;cms, B=25\;G, V=600V$ while keeping all other parameters the same and observed similar asymmetry effects. A uniform temperature of $300~\mathrm{K}$ is assumed for the Helium gas with a typical pressure of $10~\mathrm{mTorr}$. The code calculates the time evolution of singly ionized $\mathrm{He}^+$ and electrons and takes account of electron-neutral elastic \cite{Brunger1992,Fon1981} and ionization collisions \cite{Rapp1965} as well as ion-neutral elastic \cite{Phelps1994} and charge exchange collisions \cite{Smirnov}. The metastable reactions are not considered because of the ambient low pressure. At high voltages secondary electrons generated by either ion or electron-induced emission can modify the discharge\cite{Khrabrov_PSST_24_2015}. However, for the sake of simplicity in this paper we only focus on the physics of the rf sheath and its modification by a weak magnetic field and neglect the generation of secondary electrons. The more complex interaction with secondary electrons will be discussed in future publications. The initial ion and electron temperatures are taken to be  $0.03~\mathrm{eV}$ and $2.5~\mathrm{eV}$ respectively. The number of cells is taken to be $3403$ and the cell size is therefore $10~\mathrm{cm}/3403 \sim 2.93 \times 10^{-3} \mathrm{cm} $ which is sufficient to resolve the electron Larmor radius of $0.22~\mathrm{cm}$ at the highest magnetic field of $70~\mathrm{G}$. The cell size, $\Delta x$, is also sufficiently small to resolve the initial Debye length ($\lambda_{de}$) of $2.35 \times 10^{-2}~\mathrm{cm}$.  The time step for our simulations, $\Delta t$, is taken to be $7.834 \times 10^{-12}~\mathrm{s}$ to ensure that the code strictly follows the stability criterion of $\Delta x/\lambda_{de} < 0.5$ and $\omega_{pe} \Delta t < 0.2$. We take about $\sim 400$ superparticles per cell. 
  Our simulation results, discussed next, pertain to steady state values, which is verified by comparing the derivative of the ion flux with the ionization source. Anomalous transport effects \citep{Hagelaar_PPCF_53_2011, Gaboriau_APL_104_2014} which may be important for an overall two-dimensional structure of the discharge have been neglected in our  1-D simulations. Incorporating anomalous transport effects in a 1D simulation model is difficult without introducing artificial (non-consistent) mechanisms. Hence for simplicity we have restricted ourselves to considering only collisional transport effects. 
 We do not solve the neutral gas dynamics and therefore the background gas is uniformly distributed maintaining the gas pressure at $10~\mathrm{mTorr}$. The electrodes have perfectly absorbing boundary conditions. We would also like to mention that our simulations do not take into account the matching network and blocking
capacitor so that the self-bias of powered electrode is
assumed to be zero. 

\begin{figure}
\centering
\includegraphics[width=0.5\textwidth]{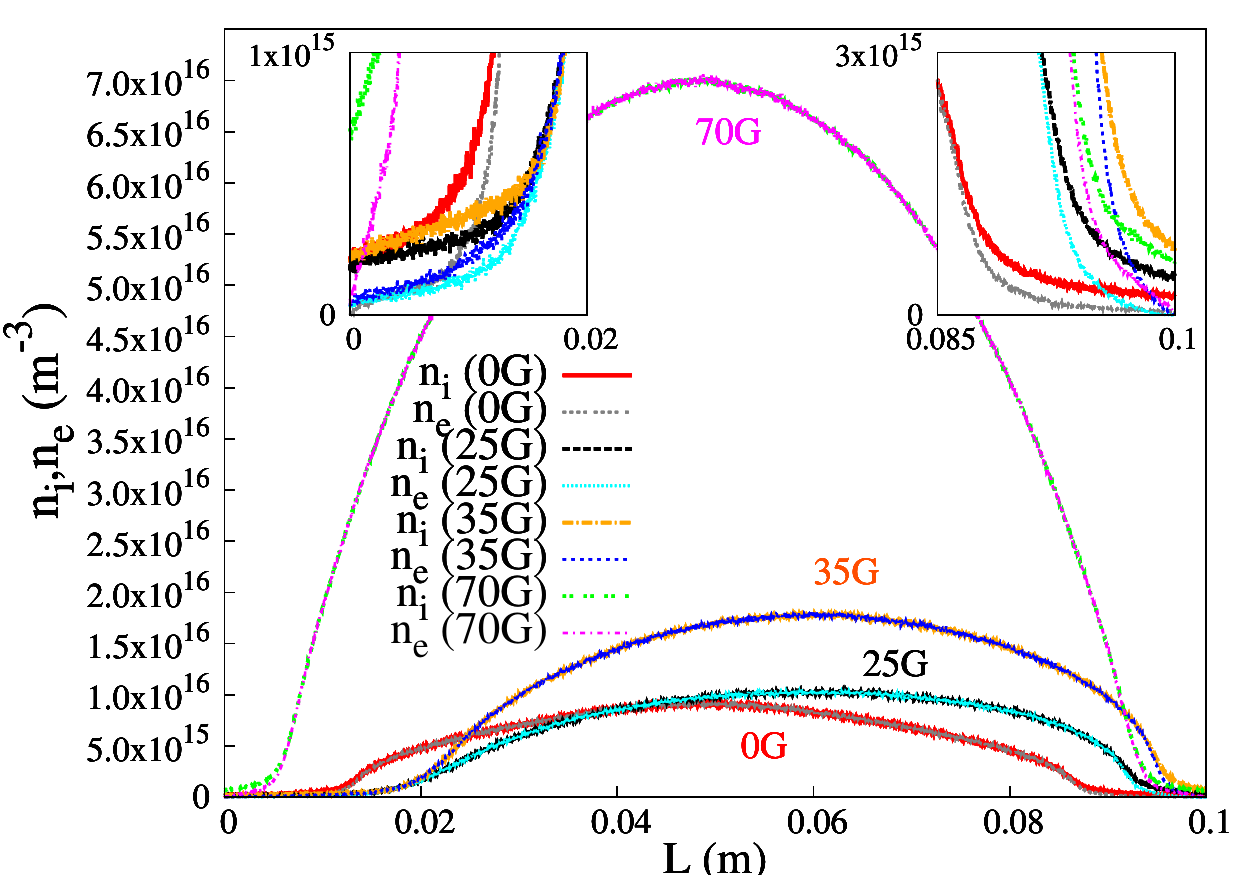}
 \caption{Variation of the time averaged spatial profiles of $n_e$ and $n_i$ with changes in the applied magnetic field B. The insets show details of the profiles near the electrodes and highlight the asymmetry of the sheath structures. The asymmetry is maximum at $35~\mathrm{G}$ for fixed values of other parameters of the simulation.
 Here the PE and GE are at $0~\mathrm{cm}$ and $10~\mathrm{cm}$ respectively}.
 \label{Fig1}
\end{figure}

We now discuss our simulation results. 
Fig~(\ref{Fig1}) shows the spatial profiles of the plasma electron and ion densities, $n_e$ and $n_i$ respectively, in the gap between the 
electrodes for various values of $B$. Here, the powered electrode (PE) is at $0~\mathrm{cm}$ and the grounded electrode (GE) 
is at $10~\mathrm{cm}$. For $B=0~\mathrm{G}$, the peak bulk density of $\sim$ $9.2\times10^{15}$ $\mathrm{m}^{-3}$ is located at 
the center of the discharge ($5.0~\mathrm{cm}$) and the two electrode sheaths are identical with widths of
$1.5~\mathrm{cm}$. The sheath width is taken to be the maximum distance from the electrode where the quasi-neutrality condition breaks down\cite{LibLicPPDM}. The ion and electron densities, $n_i$, $n_e$, at 
the left and right electrodes are $2.0\times10^{14}$ $\mathrm{m}^{-3}$ and $8.0\times10^{13}$ $\mathrm{m}^{-3}$ 
respectively with $n_e<n_i$. By increasing the strength of $B$ (e.g. to $30~\mathrm{G}$), 
the peak value of the bulk plasma density shifts towards the GE to $6.2~\mathrm{cm}$
and the density of the bulk plasma also increases up to 
$1.33\times10^{16}$ $\mathrm{m}^{-3}$ at $30~\mathrm{G}$. An increase in the bulk density implies an increase in the ion flux.
We also note that the sheath width near GE is narrow ($\sim0.75~\mathrm{cm}$) compared to the
sheath width near the PE ($\sim1.9~\mathrm{cm}$) [see insets]. So by controlling the sheath width (through $B$) we can control the 
potential drop across it and hence the ion energy. We also find that $n_i$ 
at PE ($2.1\times10^{14}$ $\mathrm{m}^{-3}$) is nearly 3 times less than that at GE 
($6.25\times10^{14}$ $\mathrm{m}^{-3}$) and $n_e>n_i$ at 
PE ($n_e$ $\sim$ $2.6\times10^{14}$ $\mathrm{m}^{-3}$) during short interval of an RF cycle. At GE $n_i>n_e$ but curiously $n_e - n_i$ does not go to a minimum at the end of the RF cycle which is at variance from the behavior in a normal single frequency CCP discharge.
At $B=35~\mathrm{G}$, the sheath width near GE is $0.6~\mathrm{cm}$ and at PE it is $1.71~\mathrm{cm}$. 
The center of the bulk plasma is at $\sim6.2~\mathrm{cm}$. Here $n_i$ at GE ($\sim$ $8.0\times10^{14}$ $\mathrm{m}^{-3}$) is more than 3 
times higher compared to PE ($\sim$ $2.5\times10^{14}$ $\mathrm{m}^{-3}$). Again like the $30~\mathrm{G}$ case, 
$n_e$($\sim$ $3.1\times10^{14}$ $\mathrm{m}^{-3}$) at PE is higher than $n_i$ 
($\sim$ $2.5\times10^{14}$ $\mathrm{m}^{-3}$) and  at GE ($\sim$ $3.0\times10^{13}$ $\mathrm{m}^{-3}$) 
is nearly 25 times lower than $n_i$ ($\sim$ $8.0\times10^{14}$ $\mathrm{m}^{-3}$), which is a very
unusual phenomenon since in general $n_i>n_e$ at the electrodes in normal CCP discharges. 
This unusual phenomenon is attributed to the strange shape of the potential profile existing over a short duration
of an RF period. We will discuss the physical reason of this abnormal 
behavior later in the text. By increasing $B$ (from $40~\mathrm{G}$ to $70~\mathrm{G}$) the density of the bulk plasma increases 
and the center of the peak density shifts towards PE. Finally at $70~\mathrm{G}$, the center of 
peak density is at $5~\mathrm{cm}$ and both sheaths are again nearly symmetric with a sheath width 
of $\sim0.8~\mathrm{cm}$. The ion density at both the right and the left electrode is 
$\sim6.5\times10^{14}~\mathrm{m}^{-3}$ and the electron density is 
$\sim 2.0\times10^{14}$ $\mathrm{m}^{-3}$ i.e. $n_i>n_e$ which is similar to the $B=0~\mathrm{G}$ case. The 
sheath width at $70~\mathrm{G}$ is less compared to $0~\mathrm{G}$ ($\sim 1.5~\mathrm{cm}$) because the density in the
former case ($\sim$ $7.0\times10^{16}$ $\mathrm{m}^{-3}$) is nearly 8 times high compared to 
latter case ($\sim$ $9.0\times10^{15}$ $\mathrm{m}^{-3}$).
\begin{figure}
\includegraphics[width=0.5\textwidth]{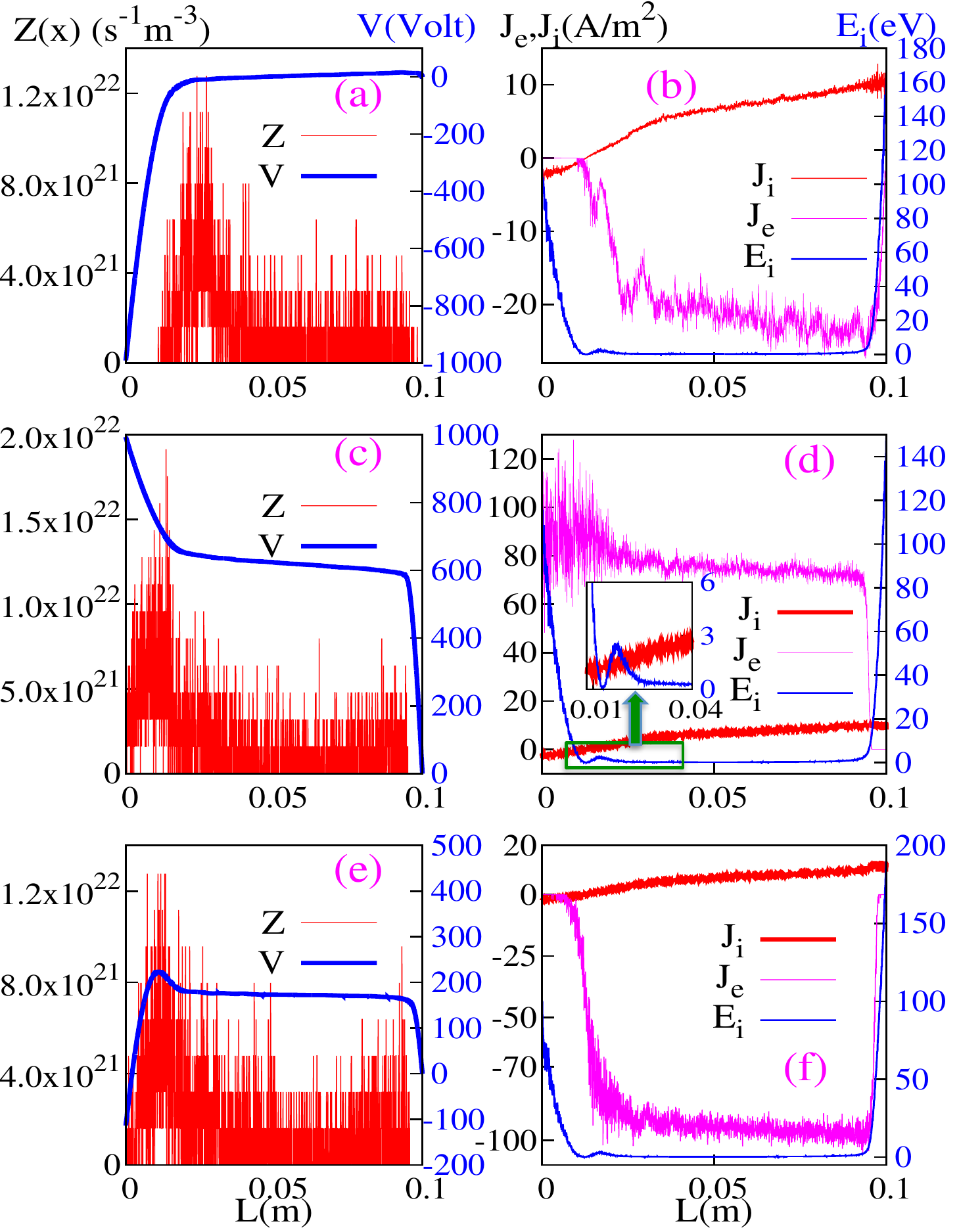}
 \caption{Spatial profiles of potential ($V$), rate of volume ionization ($Z(x)$), ion current density ($J_i$), electron current density ($J_e$) and ion energy ($E_i$) for three different phases of $V$ during an RF cycle for $B=35~\mathrm{G}$. The inverted potential profile seen in (c) confines the ions and accelerates the electrons.
These accelerated electrons create an additional volume ionization peak seen in the Z(x) curves. The ion current $J_{i}$ spatial profiles show a distinct asymmetry between PE and GE. }
 \label{Fig2}
\end{figure}

Fig ~(\ref{Fig2}) shows the rate of volume ionization ($Z(x)$), ion current density ($J_i$), 
electron current density ($J_e$) and ion energy ($E_i$) corresponding to three different 
phases (i.e. 0, $\pi$ and $3\pi/2$) of the applied potential ($V$) during an RF cycle for the $35~\mathrm{G}$ case. Here, the time averaged ion energy (local, per particle) in steady state is about $\sim2.1~\mathrm{eV}$ and the corresponding ion Larmor radius is  $\sim 8.5~\mathrm{cm}$.
In this figure, 
the values of $V$ at the powered electrode for the panels ((a), (b)), ((c), (d)) and ((e), (f)) 
are $-1000~\mathrm{V}$, $1000~\mathrm{V}$ and $-100~\mathrm{V}$ respectively. $Z(x)$ is nearly uniform at $0~\mathrm{G}$, but becomes nonuniform and shifts towards the powered 
electrode when $B$ is increased. The magnitude of ionization rate $Z(x)$ is maximal, i.e. $\sim$ $1.9\times10^{22}$ $\mathrm{s}^{-1}~\mathrm{m}^{-3}$ for $35~\mathrm{G}$ at $x\approx 1.3~\mathrm{cm}$. This phenomenon can be understood by looking at the 
profile of potential ($V$) in Fig~(\ref{Fig2}) (c). Here, the potential difference between 
the PE ($x=0~\mathrm{cm}$) and the center of bulk plasma ($i.e.$ $V_{PB}=V_{PE}-V_{BP}$) is $\sim 380~\mathrm{V}$ 
and the distance over which this potential difference exists is 0 to 3 cm. A similar type of potential 
profile was also reported by Kushner \cite{JAP_2003_94_1436}. Here,  
$V_{PE}$ and $V_{BP}$ are the potentials at PE and the center of discharge respectively. 
For $B=0~\mathrm{G}$, the electrons become lost to the electrode and make it 
negatively charged leading to the development of a positive space charge ion sheath near the electrode. 
The potential in the bulk plasma is always higher compared to the PE. The potential drops 
across the sheaths accelerate the ions towards the electrodes and confine the electrons 
inside the bulk plasma. In general, the ions are lost continually and in equal amounts at both sheaths 
but the electrons are lost at both sheaths during only a small fraction of an 
RF period, namely, when the repelling-electrons electric field at the wall reaches its minimum 
(such as in the $0~\mathrm{G}$ case). However, when the magnitude of $B$ is significant (i.e. $35~\mathrm{G}$ here) 
so that the electrons are magnetized  while the ions are not, an inverted potential profile Fig~(\ref{Fig2}) (c) is developed between the PE and the center of the discharge ($V_{PB}$). Such a phenomenon can be understood as follows.
For $B=0~\mathrm{G}$ case, since the ions are lost continuously and in equal amounts at both sheaths 
the ion current density $J_i$ is symmetric and equal at both electrodes ($\sim 4~\mathrm{A}/\mathrm{m}^2$). 
For $B=35~\mathrm{G}$, the electrons do not get a chance to be lost at the GE (in an RF cycle) 
as the electrode sheath never collapses (the time when electrons are lost to the electrode). 
Under these conditions the electrons can only be lost from the PE when the biased plate 
is positive with respect to the plasma (as shown in Fig~(\ref{Fig2})(c)). Because there is a 
continuous loss of ions at the GE in order to conserve current, most of the electrons, which 
are formed due to strong ionization near the PE, are absorbed at the PE in a fraction of an RF 
cycle. On the other hand, the ions are pushed 
towards the GE by the potential $V_{PB}$. The ion loss rate is thus mostly towards the GE and exhibits a strong asymmetry. 
By observing the profile of $J_i$ it is clear that the ion current on the left 
electrode is ($\sim2~\mathrm{A}/\mathrm{m}^2$) and is much lower than the right electrode ($\sim11~\mathrm{A}/\mathrm{m}^2$).
It is important to note here that the potential $V_{PB}$ accelerates the electrons 
and stops the ions when the charged particles move towards the PE. These magnetized electrons 
accelerate under $V_{PB}$ creating a large rate of ionization near $\sim 1.25~\mathrm{cm}$ 
which is nearly 4 times higher than the $0~\mathrm{G}$ case. The PE absorbs the electrons generated 
due to ionization and the ions are pushed back towards the GE. The ion acceleration can be 
identified in Fig~(\ref{Fig2}) where the ion energy $E_i$ is seen to increase from $\sim 1.25~\mathrm{cm}$ to $2.2~\mathrm{cm}$ which is higher than the $E_i$ from $5.0~\mathrm{cm}$ to $\sim 9.0~\mathrm{cm}$ (up to the sheath edge near the GE).

We have also measured the time evolution of electric field, ($E~\mathrm{(V/m)}$), at both the PE and the GE 
for $0~\mathrm{G}$ and $35~\mathrm{G}$ cases. At $0~\mathrm{G}$, only the magnitude of $E$ changes at 
both electrodes (due to expansion and collapse of electron sheath). Conversely at $35~\mathrm{G}$, 
the electric field at GE just changes its magnitude (like $0~\mathrm{G}$) but at the PE $E$ (magnitude of $E$ at PE is much higher compared to $E$ at GE) not only changes its magnitude but also changes its direction for a short interval of time (due to $V_{PB}$). 

\begin{figure}
\includegraphics[width=0.5\textwidth]{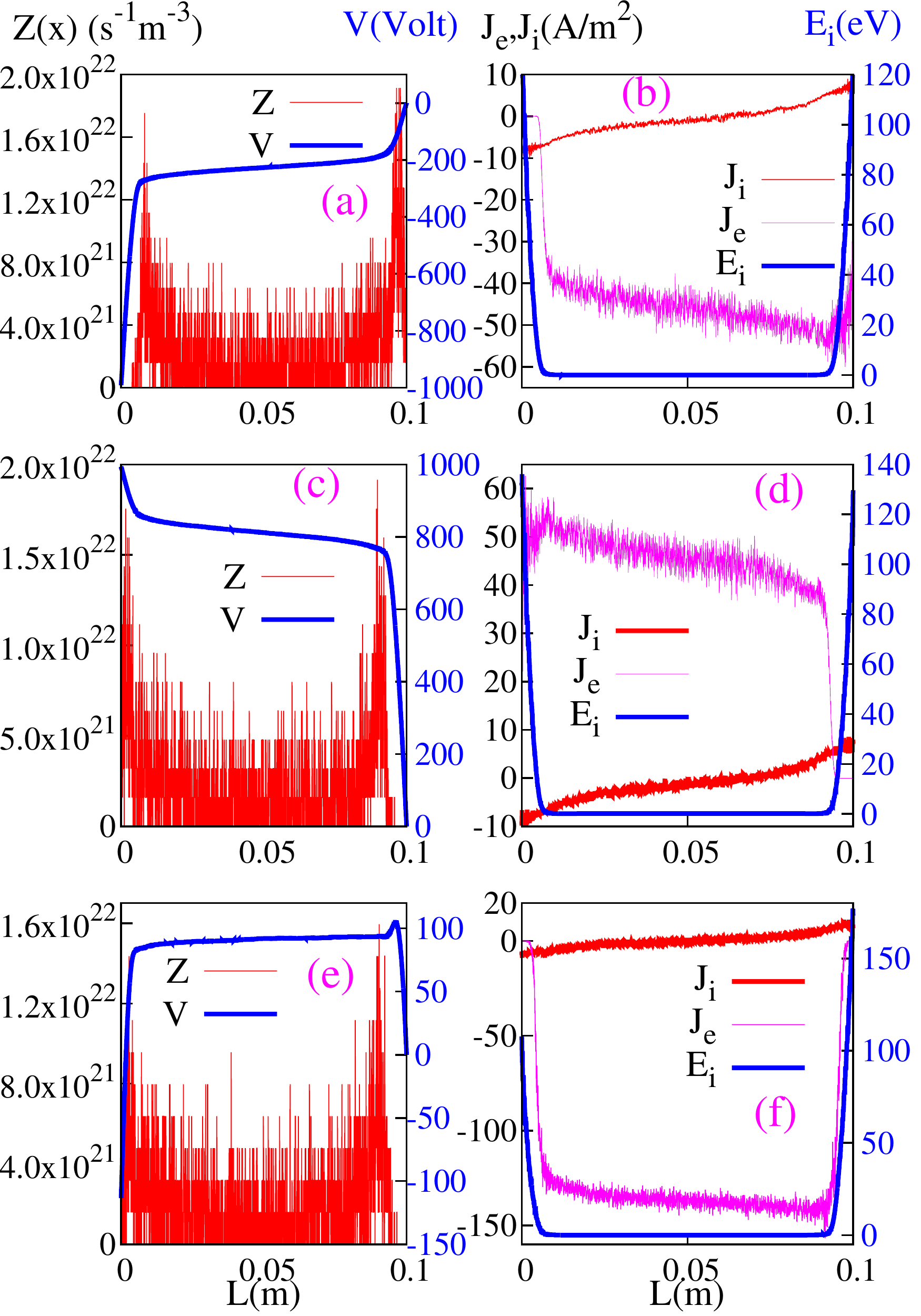}
 \caption{Spatial profiles of $V$, $Z(x)$, $J_i$, $J_e$ and $E_i$ for three different phases of $V$ during an RF cycle for $B=70~\mathrm{G}$. At this higher value of the magnetic field the magnitude of the inverted potential profile decreases and the ionization profile starts to become symmetric again with respect to the center of discharge. The ion current profile also becomes symmetric.}
 \label{Fig3}
\end{figure}

The physical consequences of the changed dynamics of electrons and ions continue to be interesting 
when $B$ is further increased to $70~\mathrm{G}$. Here, electrons are strongly magnetized and ions are
weakly magnetized. As shown in Fig~(\ref{Fig3}), the magnitude 
of $Z(x)$ is almost symmetric between both sides from center of discharge. It is maximum near the sheath edges ($\sim$ $1.6\times10^{22}$ $\mathrm{s}^{-1}\mathrm{m}^{-3}$) and is nearly uniform ($\sim$ $6.0\times 10^{21}$ $\mathrm{s}^{-1}\mathrm{m}^{-3}$) elsewhere. Again this phenomenon 
can be understood by observing the profile of the potential (V) in Fig~(\ref{Fig3}) (c). Here, the 
potential difference between the PE and the center of bulk plasma 
($i.e.$ $V_{PB}=V_{PE}-V_{BP}$) is $\sim130~\mathrm{V}$, which is nearly 3 times 
less compared to $B$=$35~\mathrm{G}$. Furthermore, the length in which this potential difference 
has been developed is $0.65~\mathrm{cm}$ that is nearly 5 times less compared to $35~\mathrm{G}$ case. 
So it is clear that for $B=70~\mathrm{G}$, the electrons are not only accelerated by a smaller 
$V_{PB}$ but also for a very short distance ($0.65~\mathrm{cm}$) near to both electrodes which makes the 
ionization profile symmetric w.r.t. center of bulk plasma. The magnitude of $Z(x)$ is  also lower compared to $35~\mathrm{G}$ case. Also this $V_{PB}$ at $70~\mathrm{G}$ is not strong 
enough to stop the majority of ions when they move towards the PE. A curious feature, observed for the high magnetic field case and that arises from the changed mobility of the electrons, is the development of an inverse sheath in front of both the PE and the GE. This can be clearly seen from Fig.~3(a) and (c) and is a phenomenon that needs to be explored in more depth in the future.

As remarked earlier, for $B=0~\mathrm{G}$, the potential in the bulk plasma is always higher 
(or positive w.r.t. the ground) than that at GE. However, it is surprising that for $B=70~\mathrm{G}$ the 
bulk potential is lower (i.e. negative) w.r.t. the GE for a short time when $V$ at the powered 
electrode is $-1000~\mathrm{V}$ (see Fig~(\ref{Fig3})(a)). The potential $V$ at the PE is also slightly higher than the bulk 
potential when $V$ approaches $1000~\mathrm{V}$. This is indicative of an interesting result, namely, 
the existence of an electron-rich sheath near the GE and PE during a short interval of an RF period.

For the $B=0~\mathrm{G}$ case, the ions are lost equally and continually from both the sheaths 
 but the electrons are lost from both the sheaths during a fraction of an RF 
period when the electron sheath edge reaches a minimum distance from the electrodes. 
However at $70~\mathrm{G}$, the potentials $V_{PB}$ at the powered and $V_{GB}$ at the grounded electrode 
accelerate the electrons and have a negligible effect on the ion motion when the charge particles move 
towards the electrodes. Here, $V_{GB}$ is the potential difference between the GE and the bulk plasma 
(see Fig~(\ref{Fig3})(a)). The ions are lost at both electrodes continuously, however, electrons are 
collected by both PE and GE alternatively during short intervals of an RF cycle. 

Like in the unmagnetized case, at $70~\mathrm{G}$, $J_i$ is symmetric and equal at both electrodes ($\sim7~\mathrm{A}/\mathrm{m}^2$).  
The GE and PE are positive 
(see Fig~(\ref{Fig3}) (a) and (c)) w.r.t. the bulk plasma during a small interval of an RF cycle. During this 
particular phase of an RF cycle, the electrons are pulled in by both the electrodes 
alternatively due to $V_{GB}$ and $V_{PB}$ (see profile of $J_e$ in Fig~(\ref{Fig3})(b) and (d)). 
The profile of ion energy $E_i$ (Fig~(\ref{Fig3}) (b,d,f)) shows that the ions flow smoothly as in the
unmagnetized case.

We have also measured the time evolution of $E~\mathrm{(V/m)}$ at both PE and 
GE for $B=70~\mathrm{G}$ and found that it not only changes 
its magnitude but also changes its direction for short intervals of time 
(due to $V_{GB}$ and $V_{PB}$).

\begin{figure}
\includegraphics[width=0.5\textwidth]{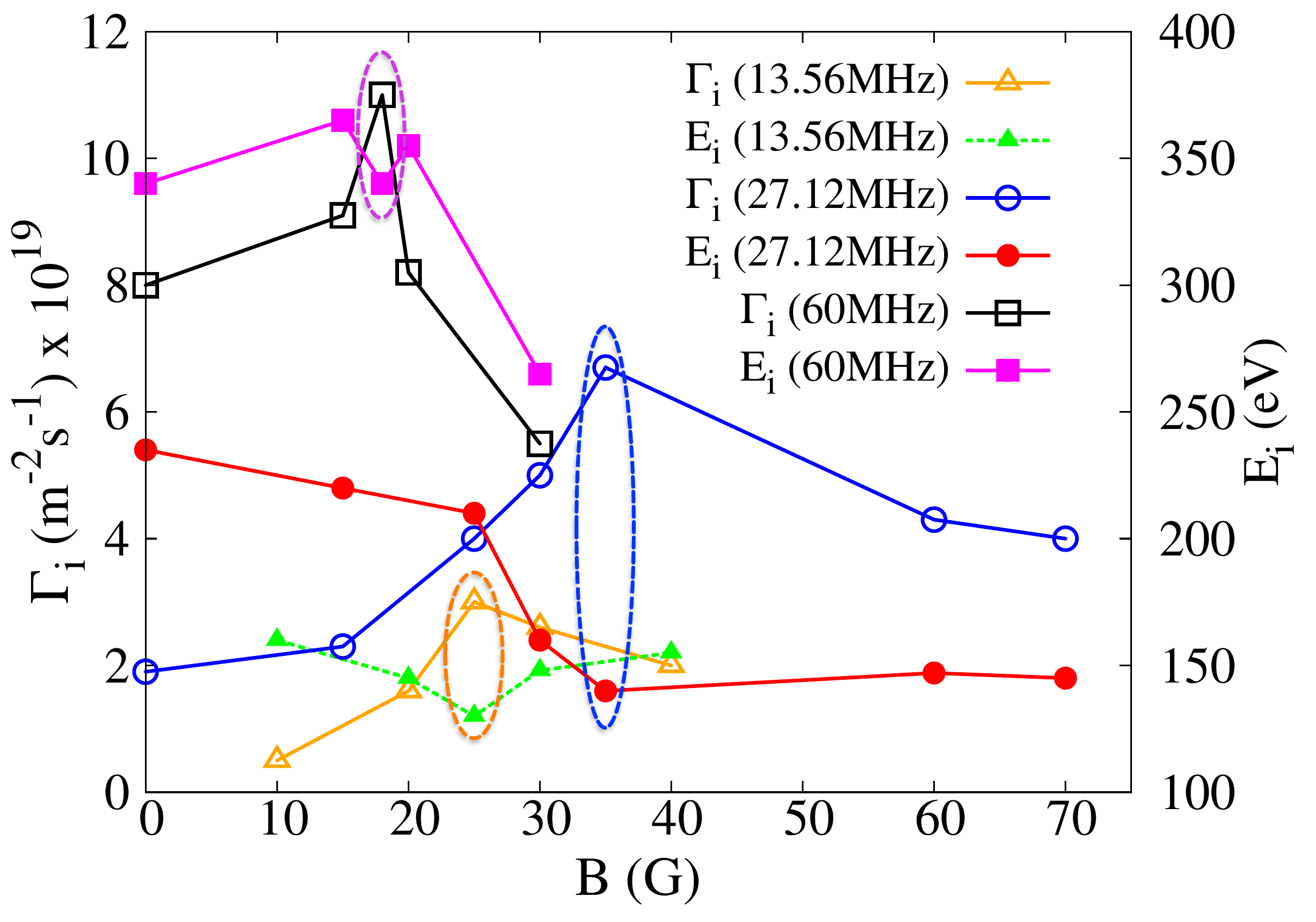}
 \caption{Variation of $\Gamma_i$ and $E_i$ at the grounded electrode as a function of B for 3 different applied frequencies.}
 \label{Fig4}
\end{figure}

It is important to point out at this stage
that the asymmetry effect in the sheaths can also be influenced by other 
operating parameters of the device, such as the applied frequency, the pressure, the applied voltage 
and the type of gas. To highlight one such influence we have studied the applied frequency effect 
by repeating our simulations at different values of $B$ over a range of frequencies, namely, 
$13.56~\mathrm{MHz}$, $27.12~\mathrm{MHz}$ and $60~\mathrm{MHz}$.
The changes in the the ion flux ($\Gamma_i$) and in the ion energy  ($E_i$) at GE for three different 
frequencies are shown in Fig~(\ref{Fig4}). The dotted ellipticals are drawn to highlight the fact that at a particular frequency one can get maximum ion flux with minimum ion energy for an appropriate choice of the magnetic field strength. Here, $\Gamma_i$ at $13.56~\mathrm{MHz}$ is small compared 
to the ion fluxes at other frequencies. We see that for chosen 
specific values of $B$ it is possible to get a maximum in $\Gamma_i$ and a minimum in $E_i$ simultaneously in  each case. For $13.56~\mathrm{MHz}$ at $25~\mathrm{G}$ the maximum of $\Gamma_i$ is 
$3.0\times10^{19}$ $\mathrm{m}^{-2}\mathrm{s}^{-1}$ and the minimum  $E_i$ is $130~\mathrm{eV}$. 
Similarly for $27.12 MHz$ (at $35~\mathrm{G}$) and $60~\mathrm{MHz}$ (at $18~\mathrm{G}$) the maximum values of
$\Gamma_i$ are $6.7\times10^{19}$ $\mathrm{m}^{-2}s^{-1}$ and $1.1\times10^{20}$ $\mathrm{m}^{-2}\mathrm{s}^{-1}$ 
respectively while the minimum values of $E_i$ are $140~\mathrm{eV}$ and 
$342~\mathrm{eV}$ respectively. Likewise, variations in other basic parameters such as pressure, applied voltage or
type of gas in the presence of a magnetic field reveal a rich operating space for the SF-CCP where
high ion flux with a simultaneous control of ion energy can be achieved. Details of these additional simulations will be reported in follow up publications.

In summary, our simulations provide a proof-of-principle demonstration of the effective 
use of a static magnetic field to significantly improve the operating characteristics of 
a standard CCP and thereby provides a simple alternative technique to simultaneously 
control the ion flux and ion energy in such devices. The basic physical mechanism underpinning 
this control is the altered dynamics of the electrons and ions under the influence of the magnetic 
field that impacts the location and magnitude of the ionization region as well as the width of the
sheaths at the electrodes. The magnitude of the magnetic field is chosen to be such that the 
electrons are strongly magnetized resulting in their reduced mobility across the magnetic field
while the ions remain relatively unmagnetized. It is possible 
then to optimize the operating parameters of the device in a desired manner by a suitable choice of
the other basic parameters of the device. Our simulations have been carried out for physically realistic
values of plasma parameters (a low pressure He discharge) and a magnetic field of $\sim35~\mathrm{G}$ that can be
easily recreated in an laboratory set up to experimentally test the basic concept. In an actual experiment deviations from our simulation results are likely to arise due to the limitations of our model 
calculations such as the one dimensional approximation, unavoidable non-uniformity in the magnetic field, assumption of a constant voltage operation rather than a constant power operation, etc. However, it is hoped that our present findings can become the basis for both further numerical simulations (2d and 3d) 
as well as experimental explorations of this concept in order to assess the feasibility of a practical device exploiting
this technique.\\

The authors would like to thank Shali Yang for her help in cross checking of some of our simulation results  on another PIC/MCC code (iPM). 

\bibliographystyle{unsrt}

\end{document}